\documentclass[prb,twocolumn,amsmath,amssymb]{revtex4}

\usepackage{graphicx}
\usepackage{dcolumn}
\usepackage{bm}

\begin{document}


\title{Effective action for Bose-Einstein condensates}

\author{Takafumi Kita}
\affiliation{Department of Physics, Hokkaido University, Sapporo 060-0810, Japan}%

\date{\today}

\begin{abstract}
We clarify basic properties of an effective action (i.e., self-consistent perturbation expansion) for interacting Bose-Einstein condensates,
where field $\psi$ itself acquires a finite thermodynamic average $\langle \psi\rangle$ besides two-point Green's function $\hat G$  to form an off-diagonal long-range order. 
It is shown that the action can be expressed concisely order by order in terms of the  interaction vertex and a special combination of $\langle\psi\rangle$ and $\hat G$
so as to satisfy both Noether's theorem and Goldstone's theorem (I) corresponding to the first proof. 
The self-energy is predicted to have a one-particle-reducible structure due to $\langle \psi\rangle\neq 0$ to 
transform the Bogoliubov mode into a bubbling mode with a substantial decay rate.
\end{abstract}

\maketitle

\section{Introduction}

Self-consistent approximations have played a crucial role for clarifying basic or exotic properties of diverse condensed matter systems.
Lowest-order ones with respect to the interaction are generally known as mean-field or molecular-field theories,
which already include many outstanding examples such as the Hartree-Fock approximation for normal states, Weiss and Stoner theories for ferromagnetism, and
Bardeen-Cooper-Schrieffer theory for superconductivity.
In general, the self-consistent scheme has a notable advantage over the simple perturbation expansion that spontaneous symmetry-breaking phases can be described on an equal footing.

This approach can be  improved systematically to include higher-order correlations
based on a self-consistent perturbation expansion with Matsubara Green's function,\cite{LW60,dDM64,BS89,Kita09,Kita11b} 
as first shown by Luttinger and Ward for normal states.\cite{LW60}
Another advantage of this method is that it satisfies various conservation laws, i.e., Noether's theorem, \cite{Weinberg96}
as pioneered by Kadanoff and Baym\cite{KB62}  and later shown unambiguously by Baym.\cite{Baym62} 
Hence, it can be used to describe nonequilibrium phenomena, including their approach to equilibrium,
by transforming the imaginary-time Matsubara contour into the real-time Schwinger-Keldysh contour. \cite{Schwinger61,Keldysh64,Kita10}
Later, this  ``$\Phi$-derivable'' or ``conserving'' approximation scheme has
also been generalized to relativistic quantum field theory 
by Cornwall, Jackiw, and Tomboulis (CJT) \cite{CJT74}
to find a wide field of applications in high-energy physics
with the name of ``two-particle-irreducible (2PI) effective action.''\cite{KIV01,Berges02}

However, extending this powerful formalism to Bose-Einstein condensates (BECs) has encountered
a serious  ``conserving vs.\ gapless'' dilemma that either
Noether's theorem for conservation laws or Goldstone's theorem (I) for spontaneously broken symmetries is violated 
in standard approximations,\cite{Weinberg96,GSW62,JL64}
as first pointed out by Hohenberg and Martin in 1965.\cite{HM65,Griffin96}
The basic difficulty lies in how to renormalize the condensate wave function $\Psi\equiv \langle \psi\rangle$ and quasiparticle Green's function $\hat G$ consistently 
in the presence of  tadpole and other anomalous interaction vertices characteristic of BECs (see Fig.\ \ref{Fig2}(b)-(d) below)
so as to satisfy the two fundamental theorems simultaneously.

It was shown in a previous paper\cite{Kita09} that this long-standing problem may be resolved successfully by
extending the Luttinger-Ward theory to BECs with the help of an identity for the interaction energy.
The resultant formalism reveals that there should be a new class of Feynman diagrams for
the self-energy that has been overlooked so far, i.e., those that may be classified as 
``one-particle reducible'' (1PR) due to tadpole and other anomalous interaction vertices.\cite{Kita11}
The rationale for their existence is that they are indispensable for the identity to be satisfied
order by order in the self-consistent perturbation expansion with respect to the interaction in the same way as  the Luttinger-Ward functional.
This novel  structure of the self-energy has been predicted to modify standard results based on the Bogoliubov theory\cite{Bogoliubov47,LHY57,AGD63,GN64,SK74,WG74,Griffin93} substantially.
For example, we have pointed out in a previous paper\cite{TK13} that it will add a term to the Lee-Huang-Yang expressions\cite{LHY57} 
for the ground-state energy  and condensate density of the dilute Bose gas; see Eq.\ (\ref{LHY-E}) below for further details.
We have also shown\cite{TK14} that it will transform the single-particle Bogoliubov mode with an infinite lifetime into a bubbling mode with a considerable decay rate that is
proportional to the $s$-wave scattering $a$ in the dilute limit.
Finally,  this single-particle bubbling mode  should be different in character from the two-particle collective excitation,\cite{Kita10b,Kita11}
contrary to the standard understanding where both are considered identical.\cite{GN64,SK74,WG74,Griffin93}
Nevertheless, the two modes may be regarded separately as Nambu-Goldstone bosons corresponding to the two different proofs;\cite{GSW62,Weinberg96}
their contents should be distinguished clearly as ``Goldstone's theorem (I)'' and ``Goldstone's theorem (II)''\cite{Kita11}
with the former being identical to the Hugenholtz-Pines theorem.\cite{HP59}

Now, purposes of the present paper are twofold. First, we report a further refinement of this formalism,
whose key quantity has been the $\Phi$ functional, $\Phi=\Phi[\Psi,\bar\Psi,\hat G]$, given as a power series of the interaction.\cite{Kita09}
We will show that it can be transformed into a functional $\Phi=\Phi[\hat{\cal G}]$ of a single quantity $\hat{\cal G}$
that is defined in terms of $\hat G$, $\Psi$, and $\bar\Psi$ as Eq.\ (\ref{calG}) below.
Thus, the condensate wave function apparently disappears from $\Phi$,
thereby resulting in a considerable reduction in the number of Feynman diagrams to be considered.
This fact will be checked specifically up to the forth order of the expansion with respect to the interaction.
It will also be exemplified that any single diagram of the normal state can be a source of an approximate $\Phi$ for BECs
that satisfies both Noether's theorem and Goldstone's theorem (I).
Second, we will trace possible origins of a discrepancy between the present 
$\Phi$ with a one-particle-irreducible (1PI) structure and 
the one given by CJT,\cite{CJT74} which consists of 2PI diagrams 
that cannot be separated by cutting any pair of $\hat G$ lines even 
for spontaneous symmetry-breaking phases of $\langle \psi\rangle\neq 0$. 

In Sec.\ \ref{sec2}, we transform results of the previous study\cite{Kita09} 
into the Lagrangean formalism with path integrals.
The discrepancy between the CJT \cite{CJT74} and present formalisms 
is discussed in Sec.\ \ref{subsec:CJT}  in the context of Bose-Einstein condensation.
In Sec.\ \ref{sec3}, it will be shown that the $\Phi$ functional can be given concisely as a functional of $\hat{\cal G}$
defined by Eq.\ (\ref{calG}) below.
Section \ref{sec4} provides a brief summary.

\section{Summary of previous results\label{sec2}}

\subsection{System and basic quantities}
We consider an ensemble of identical bosons with mass $m$ and spin $0$ 
described by the action \cite{Weinberg96,AS10}
\begin{equation}
S=S_0+S_{\rm int},
\label{S}
\end{equation}
with 
\begin{subequations}
\label{S_0-S_int}
\begin{align}
S_0=&\,\int   d^4 x\,\bar \psi(x)\!\left(\frac{\partial}{\partial\tau}-\frac{\hbar^2{\bm\nabla^2}}{2m} -\mu\right)\!\psi(x) ,
\label{S_0}
\\
S_{\rm int}=&\,\frac{1}{2}\int  d^4 x\int  d^4 y\,
V(|x-y|)\bar\psi(x)\bar\psi(y)\psi(y)\psi(x).
\label{S_int}
\end{align}
\end{subequations}
Here $\psi$ is the complex bosonic field and $\bar\psi$ its conjugate,
$x\equiv ({\bm r},\tau)$ specifies a space-``time'' point with $0\!\leq\! \tau \!\leq\! \beta\equiv (k_{\rm B}T)^{-1}$
($k_{\rm B}$:  Boltzmann constant,  $T$: temperature), $\mu$ is the chemical potential,
and $V$ is an interaction potential.

It is convenient to regard $\psi$ and $\bar\psi$ as elements of a column or row vector as 
\begin{equation}
\begin{bmatrix}
\vspace{1mm}
\psi \\ \bar\psi
\end{bmatrix}
\equiv \begin{bmatrix}
\vspace{1mm}
\psi_1 \\ \psi_2
\end{bmatrix}\equiv \vec{\psi},
\hspace{5mm}
\begin{bmatrix}
\bar\psi \!&\! \psi 
\end{bmatrix}
\equiv \begin{bmatrix}
\psi^1 \!&\! \psi^2 
\end{bmatrix}\equiv \vec{\psi}^{\,\dagger} ,
\end{equation}
so that $\psi^i=\psi_{3-i}$  ($i=1,2$).
Next, we define the condensate wave function $\Psi_i$ and a $2\times 2$ matrix Green's function $\hat G\equiv (G_{ij})$ in the Nambu space 
by 
\begin{subequations}
\label{GPsi-def}
\begin{equation}
\Psi_i(x)\equiv \langle \psi_i(x)\rangle,
\label{Psi-def}
\end{equation}
\begin{equation}
G_{ij}(x,y)\equiv\bigl[\langle T_\tau \psi_i(x)\psi^j(y)\rangle -  \Psi_i(x)\Psi^j(y)\bigr](-1)^j ,
\label{G-def}
\end{equation}
\end{subequations}
which obeys $G_{ij}(x_1,x_2)\!=\!(-1)^{i+j-1}G_{3-j,3-i}(x_2,x_1)\!=\!(-1)^{i+j}G_{ji}^{*}({\bm r}_2\tau_1,{\bm r}_{1}\tau_2)$.\cite{Kita09,Kita10b}
With these symmetries, it is convenient for later purposes to introduce
\begin{subequations}
\label{GFbF}
\begin{align}
G(x,y') \equiv&\,\bar G(y',x)  \equiv \frac{\displaystyle G_{11}(x,y')-G_{22}(y',x)}{2},
\label{G}\\
F(x,y)\equiv&\,\frac{G_{12}(x,y)+G_{12}(y,x)}{2},
\label{F}\\
\bar F(x',y')\equiv &\,-\frac{G_{21}(x',y')+G_{21}(y',x')}{2},
\label{bF}
\end{align}
\end{subequations}
where non-primed (primed) arguments are associated with $\psi$ ($\bar\psi$).
Function $G$ is the conventional Green's function that remains finite in normal states, 
whereas $F$ and $\bar F$ are ``anomalous'' ones 
characteristic of the off-diagonal long-range order (ODLRO) \cite{Yang62}
with $\bar F(x_1',x_2')=F^*({\bm r}_2'\tau_1',{\bm r}_1'\tau_2')$.

Inverse matrix of  $\hat G=(G_{ij})$ for $V\rightarrow 0$ can be written explicitly in terms of the operators in Eq.\ (\ref{S_0}) as\cite{Kita09}
\begin{equation}
\hat G_0^{-1}(x,y)\equiv \left[-\hat\sigma_0\frac{\partial}{\partial\tau}+\hat\sigma_3\left(\frac{\hbar ^2{\bm \nabla}^2}{2m} +\mu\right)\right]\delta(x-y) ,
\label{hatG_0^-1}
\end{equation}
where $\hat{\sigma}_0$ and $\hat\sigma_3$ denote the $2\times 2$ unit matrix and  third Pauli matrix, respectively.

It is also useful to introduce a symmetrized vertex as
\begin{align}
V_{\rm s}(x'y',xy)\equiv&\,\frac{1}{2}V(|x-y|)[\delta(x',x)\delta(y',y)
\nonumber \\
&\,+\delta(x',y)\delta(y',x)] .
\label{Gamma^(0)}
\end{align}
Using it, we can express Eq.\ (\ref{S_int}) alternatively as
\begin{align}
S_{\rm int}=&\,\frac{1}{2}\int  d^4 x'\int  d^4 y'\int  d^4 x\int  d^4 y\,
V_{\rm s}(x'y',xy)
\nonumber \\
&\,\times\bar\psi(x')\bar\psi(y')\psi(y)\psi(x).
\label{S_int2}
\end{align}

\subsection{Legendre transformation}

As shown by De Dominicis and Martin,\cite{dDM64} a Legendre transformation enables us 
to establish the stationarity of the grand potential with respect 
to $\vec\Psi$ and $\hat G$  concisely and clearly.
Let us introduce the grand partition function $Z[\vec{J},\hat K]$ for action (\ref{S}) with extra source functions
$J^i(x)=J_{3-i}(x)$ and $K_{ij}(x,y)=K_{3-j,3-i}(y,x)$ by\cite{dDM64,Weinberg96,AS10}
\begin{align}
Z[\vec{J},\hat K] \equiv&\, \int D( \bar\psi,\psi) \, \exp\biggl[-S+\sum_{i} \int  d^4 x \,J^i(x)\psi_i (x) 
\nonumber\\
& +\sum_{ij} \int  d^4 x\int  d^4 y\,\psi^j (y)K_{ji}(y,x)\psi_i (x)\biggr]  ,
\label{Z}
\end{align}
which satisfies
\begin{subequations}
\begin{align}
\frac{\delta \ln Z[\vec{J},\hat K]}{\delta J^i(x)}=&\,\Psi_i(x),\\
\frac{\delta \ln Z[\vec{J},\hat K]}{\delta K_{ji}(y,x)}=&\,\langle T_\tau \psi_i(x)\psi^j(y)\rangle
\nonumber \\
=&\,G_{ij}(x,y)(-1)^j+\Psi_i(x)\Psi^j(y).
\end{align}
\end{subequations}
Subsequently, we perform a Legendre transformation from $-\ln Z[\vec{J},\hat K] $ to $\Gamma=\Gamma[\vec\Psi,\hat G]$ as
\begin{align}
\Gamma[\vec{\Psi},\hat G] \equiv&\, -\ln Z[\vec{J},\hat K] + \sum_{i} \int  d^4 x\,J^i(x) \Psi_i(x) 
\nonumber\\
&\,
+\sum_{ij} \int  d^4 x\int d^4 y\, K_{ij}(x,y)\langle T_\tau \psi_{j}(y) \,\psi^i (x)\rangle.
\label{Gamma}
\end{align}
It satisfies 
\begin{align}
\frac{\delta \Gamma[\vec{\Psi},\hat G] }{\delta \Psi_i(x)}&=J^i(x)+2\sum_j\int  d^4 y \,\Psi^j(y)K_{ji}(y,x),\nonumber\\
\frac{\delta \Gamma[\vec{\Psi},\hat G] }{\delta G_{ji}(y,x)}&=(-1)^i K_{ij}(x,y).\nonumber
\end{align}
Especially for the cases of physical interest with $\vec{J}=\vec 0$ and $\hat K=\hat 0$, 
they are reduced to the stationarity conditions
\begin{align}
\frac{\delta \Gamma[\vec{\Psi},\hat G] }{\delta G_{ji}(y,x)}=0,\hspace{10mm}\frac{\delta \Gamma[\vec{\Psi},\hat G] }{\delta \Psi_i(x)}=0.
\label{dGamma}
\end{align}
The corresponding $\Gamma$  is known as ``quantum effective action'' or simply ``effective action'' in relativistic quantum field theory.\cite{Weinberg96,CJT74}
Note also that $\Omega[\vec\Psi,\hat G]\equiv \Gamma[\vec\Psi,\hat G]/\beta$ is the grand potential of thermodynamics.
The first equality of Eq.\ (\ref{dGamma}) is exactly the stationarity condition established diagrammatically by
Luttinger and Ward for normal states.\cite{LW60}

\subsection{Exact results}

As pointed out by Jona-Lasinio,\cite{JL64} 
partition function (\ref{Z}) for $\hat K\equiv\hat 0$  and the corresponding $\Gamma=\Gamma[\vec\Psi]$ are 
useful for obtaining formally exact results for $\vec J\rightarrow \vec 0$.
First, one can show that Green's function (\ref{G-def}) obeys the Dyson-Beliaev equation\cite{JL64,Weinberg96,Kita09,Beliaev58}
\begin{subequations}
\label{DB-GP}
\begin{align}
\hat G^{-1}=\hat G_0^{-1}-\hat \Sigma ,
\label{DB-GP1}
\end{align}
where $\hat G_0^{-1}\!=\!\hat G_0^{-1}(x,y)$ is defined by Eq.\ (\ref{hatG_0^-1}), 
and $\hat \Sigma\!=\!\hat \Sigma(x,y)$ denotes the self-energy due to the interaction 
that will be specified shortly.
Subsequently, one can prove based on the gauge invariance 
that the equation for $\vec\Psi$ is also given in terms of $\hat G_0$ and $\hat \Sigma$ by \cite{Weinberg96,GSW62,JL64,Kita09}
\begin{align}
\int d^4 y \bigl[\hat G_0^{-1}(x,y)-\hat \Sigma(x,y)\bigr]\hat\sigma_3\vec\Psi(y)=\vec{0},
\nonumber
\end{align}
where $\hat \sigma_3$ originates from the  asymmetry between $\psi_1\!=\!\psi$ and $\psi_2\!=\!\bar\psi$ under the gauge transformation.
This equation may be written concisely by regarding $x$ and $y$ as matrix indices as
\begin{align}
(\hat G_0^{-1}-\hat \Sigma)\hat\sigma_3 \vec\Psi=\vec{0} .
\label{DB-GP2}
\end{align}
\end{subequations}
Equation (\ref{DB-GP2}) embodies ``Goldstone's theorem (I)'' corresponding to the first proof,\cite{GSW62,Weinberg96}
which is reduced for homogeneous systems to the Hugenholtz-Pines relation.  \cite{HP59}
Unlike their original proof,\cite{HP59} however, 
Eq.\ (\ref{DB-GP2}) has been derived
without imposing the 1PI condition on $\hat\Sigma$.\cite{Weinberg96,GSW62,JL64,Kita09}
Equation (\ref{DB-GP2}) predicts a gapless excitation spectrum for $\hat G$.
However, standard conserving approximations such as the Hartree-Fock-Bogoliubov theory fail to meet
Eq.\ (\ref{DB-GP2}), yielding an unphysical energy gap in the excitation spectrum.\cite{HM65,Griffin96}

Finally, the interaction energy $\langle S_{\rm int}\rangle$ of Eq.\ (\ref{S_int}) can also be expressed 
in terms of $\hat G$ and $\hat \Sigma$ as Eq.\ (12) of ref.\ \onlinecite{Kita09};
it reads in the present notation as
\begin{align}
\langle S_{\rm int}\rangle =&\,-\frac{1}{4}\sum_{ij} \int d^4 x\int d^4 y \,\Sigma_{ij}(x,y){\cal G}_{ji}(y,x)
\nonumber \\
\equiv &\,-\frac{1}{4}{\rm Tr}\,\hat \Sigma \,\hat {\cal G},
\label{<S_int>2}
\end{align}
where $\hat {\cal G}=({\cal G}_{ij})$ is a matrix composed of $\hat G$ and $\vec{\Psi}$ as
\begin{align}
{\cal G}_{ij}(x,y)\equiv &\,G_{ij}(x,y)+(-1)^{i}\Psi_i(x)\Psi^j(y)
\nonumber \\
=&\, \bigl[\langle T_\tau \psi_i(x)\psi^j(y)\rangle -2\Psi_i(x)\Psi^{j}(y)\delta_{j,3-i}\bigr](-1)^j .
\label{calG}
\end{align}
The second expression has been obtained by substituting Eq.\ (\ref{G-def}).
Thus, each off-diagonal element of $\hat{\cal G}$ contains an extra term $2\Psi_i(x)\Psi^{3-i}(y)(-1)^{i}$
besides $ \langle T_\tau\psi_i(x)\psi^j(y)\rangle (-1)^j$.

\subsection{Effective action}

\begin{figure}[t]
        \begin{center}
                \includegraphics[width=0.95\linewidth]{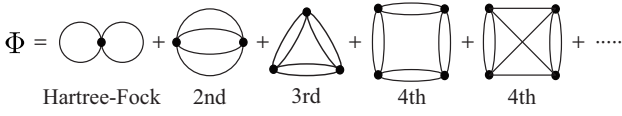}
        \end{center}
\vspace{-5mm}
        \caption{Normal-state $\Phi$ drawn without arrows. \label{Fig1}}
\end{figure}

Using $\hat G$ and $\vec\Psi$, 
we formally express $\Gamma$ of Eq.\ (\ref{Gamma}) for $\vec{J}=\vec 0$ and $\hat K=\hat 0$ in terms of 
another unknown functional $\Phi=\Phi[\vec{\Psi},\hat G]$ as \cite{Kita09}
\begin{align}
\Gamma[\vec{\Psi},\hat G] =&\, \Gamma_0-\frac{1}{2}{\rm Tr} \,\vec{\Psi}^{\dagger}\hat G_0^{-1}\hat\sigma_3 \vec\Psi
\nonumber \\
&+\frac{1}{2}{\rm Tr}\bigl[ \ln (\hat 1-\hat \Sigma \hat G_0)+\hat \Sigma\hat G\bigr]
+\beta\Phi[\vec{\Psi},\hat G]  ,
\label{Gamma-LW}
\end{align}
where $\Gamma_0$ denotes contribution of non-interacting excitations from Eq.\ (\ref{S_0}), 
and  $\hat\Sigma=\hat\Sigma[\vec\Psi,\hat G]$.
Subsequently, we perform differentiations of Eq.\ (\ref{dGamma}) 
by using Eq.\ (\ref{DB-GP}) and noting that $\vec{\Psi}^{\dagger}$ and $\vec{\Psi}$
in Eq.\ (\ref{Gamma-LW}) yield the same contribution.
Stationarity requirements of Eq.\ (\ref{dGamma}) are thereby transformed into a couple of conditions for $\Phi$ alone as
\begin{subequations}
\label{Sigma-Phi}
\begin{align}
\beta\frac{\delta \Phi}{\delta G_{ji}(y,x)}=&\, -\frac{1}{2}\Sigma_{ij}(x,y),
\label{Sigma-Phi1}
\\
\beta\frac{\delta \Phi}{\delta \Psi^{i}(x)}=&\,\sum_j \int  d^4 y\,\Sigma_{ij}(x,y)(-1)^{j-1}\Psi_j(y) .
\label{Sigma-Phi2}
\end{align}
\end{subequations}
Action (\ref{Gamma-LW}) for the normal-state limit of $(\Psi_i,G_{i,3-i})\!\rightarrow\!(0,0)$ 
is reduced to the Luttinger-Ward functional, \cite{LW60,Baym62,Kita09}
where $\Phi$ is given as a power series of $V_{\rm s}$ with
closed 2PI diagrams, i.e., those that cannot be separated by removing 
any pair of $G$ lines. It may be expressed graphically as Fig.\ \ref{Fig1},
where a filled circle denotes $V_{\rm s}$ of Eq.\ (\ref{Gamma^(0)}).

\subsection{Identities in terms of $\Phi$}

Identity (\ref{<S_int>2}) for the interaction energy can be rephrased in terms of $\Phi$.
To see this, let us replace $V\rightarrow \lambda V$ in Eq.\ (\ref{S_int}) 
and differentiate the resultant $-\ln Z[\vec{0},\hat 0;\lambda]$ from Eq.\ (\ref{Z}) with respect to $\lambda$.
Noting $(\partial/\partial\lambda) (\lambda V)=(\lambda V)/\lambda$, we obtain\cite{LW60}
\begin{subequations}
\label{dGamma/dlambda}
\begin{align}
-\frac{\partial \ln Z[\vec{0},\hat 0;\lambda]}{\partial\lambda} =\frac{\langle S_{\rm int}(\lambda)\rangle}{\lambda}
=-\frac{1}{4\lambda}{\rm Tr}\,\hat \Sigma(\lambda)\hat {\cal G}(\lambda) ,
\label{dGamma/dlambda1}
\end{align}
where we have used Eq.\ (\ref{<S_int>2}) in the second equality.
Subsequently, we replace $-\ln Z[\vec{0},\hat 0;\lambda]$ above by $\Gamma(\lambda)$ based on Eq.\ (\ref{Gamma}) and
perform its differentiation with  Eq.\ (\ref{Gamma-LW}).
Noting the stationarity conditions of Eq.\ (\ref{dGamma}), we only need to consider the explicit
$\lambda$ dependence in $\Gamma(\lambda)$ that lies in $\Phi(\lambda)$; see Fig.\ \ref{Fig1} for normal states on this point.
Thus, we also obtain
\begin{align}
-\frac{\partial \ln Z[\vec{0},\hat 0;\lambda]}{\partial\lambda}=\frac{\partial \Gamma(\lambda)}{\partial\lambda}=\beta\frac{\partial \Phi(\lambda)}{\partial\lambda}.
\label{dGamma/dlambda2}
\end{align}
\end{subequations}
Equating Eqs.\ (\ref{dGamma/dlambda1}) and (\ref{dGamma/dlambda2}) yields
\begin{align}
\beta\frac{\partial \Phi(\lambda)}{\partial \lambda}=-\frac{1}{4\lambda}{\rm Tr}\,\hat \Sigma(\lambda)\hat {\cal G}(\lambda) .
\label{<S_int>1}
\end{align}
Finally, we assume that $\Phi(\lambda)$ can be expanded from $\lambda=0$ as 
\begin{align}
\Phi(\lambda)=\sum_{n=1}^\infty \lambda^n \Phi^{(n)},
\label{Phi-exp}
\end{align}
like the Luttinger-Ward functional for normal states given graphically as Fig.\ \ref{Fig1}.
Substituting Eqs.\ (\ref{Sigma-Phi1}) and (\ref{Phi-exp}) into Eq.\ (\ref{<S_int>1}),
comparing terms of order $\lambda^{n-1}$, and setting $\lambda=1$, 
we obtain an identity for $\Phi^{(n)}$ as
\begin{subequations}
\label{identities}
\begin{align}
\Phi^{(n)}=&\,\frac{1}{2n}\sum_{ij} \int d^4 x\int d^4 y \,\frac{\delta \Phi^{(n)}}{\delta G_{ij}(x,y) }\, {\cal G}_{ij}(x,y)
\nonumber \\
\equiv &\,\frac{1}{2n}{\rm Tr}\,\frac{\delta \Phi^{(n)}}{\delta \hat G }\,\hat {\cal G}.
\label{identity1}
\end{align}
Equation (\ref{Sigma-Phi2}) is also transformed by using Eq.\ (\ref{Sigma-Phi1}) into
\begin{equation}
\frac{\delta \Phi}{\delta \Psi^{i}(x)}=-2 \sum_j \int  d^4 y\,\frac{\delta \Phi}{\delta G_{ji}(y,x)}(-1)^{j-1}\Psi_j(y) .
\label{identity2}
\end{equation}
\end{subequations}
These are the key identities corresponding to Eqs.\ (22) and (23) of ref.\ \onlinecite{Kita09} 
that have been used  to construct $\Phi$.
Indeed, the previous expressions are reproduced from Eq.\ (\ref{identities}) 
by writing $\Phi^{(n)}$ in terms of functions in Eq.\ (\ref{GFbF}) as $\Phi^{(n)}[\Psi,\bar\Psi,G,F,\bar F]$ 
and performing its differentiations.
Equation (\ref{identity1}) is thereby transformed into
\begin{subequations}
\label{identities-a}
\begin{align}
\Phi^{(n)}=&\,\frac{1}{2n}\int dx\int dy \biggl\{\frac{\delta \Phi^{(n)}}{\delta G(x,y)} [G(x,y)-\Psi(x)\bar\Psi(y)]
\nonumber \\
& +\frac{\delta \Phi^{(n)}}{\delta F(x,y)}[F(x,y)-\Psi(x)\Psi(y)]
\nonumber \\
& +\frac{\delta \Phi^{(n)}}{\delta \bar F(x,y)} [\bar F(x,y)-\bar\Psi(x)\bar \Psi(y)]\biggr\}.
\label{identity1a}
\end{align}
Equation (\ref{identity2}) for $i=1$ also reads
\begin{align}
\frac{\delta \Phi}{\delta \bar \Psi(x)}=- \int  d^4 y\,\left[\frac{\delta \Phi}{\delta G(y,x)}\Psi(y)+2\frac{\delta \Phi}{\delta \bar F(y,x)}\bar \Psi(y)\right] .
\label{identity2a}
\end{align}
\end{subequations}
These are exactly Eqs.\ (22) and (23) of ref.\ \onlinecite{Kita09}.

\begin{figure}[b]
\begin{center}
\includegraphics[width=0.9\linewidth]{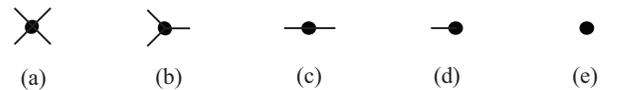}
\end{center}
\vspace{-5mm}
\caption{Diagrammatic representation of the two-body interaction in terms of the excitation field $\psi_{i}'\equiv \psi_i-\langle\psi_i\rangle$. 
Each missing line in (b)-(e) compared with (a) corresponds to $\langle\psi_i\rangle$. \label{Fig2}}
\end{figure}

It is worth pointing out that expression (\ref{Gamma-LW}) for $\Gamma$ becomes exact 
when $\Phi$ satisfies Eq.\ (\ref{identity1}) at each order up to $n=\infty$.
This can be shown in exactly the same way as for the normal state\cite{LW60}
with $V\rightarrow \lambda V$ in Eq.\ (\ref{S_int}) as follows.
First, action $\Gamma(\lambda)$ obeys
first-order differential equation (\ref{dGamma/dlambda2}).
Second, $\Gamma(\lambda)$ is identical with $-\ln Z[\vec{0},\hat 0;\lambda]$ at $\lambda=0$, i.e., for the non-interacting case.
Hence, we conclude that $\Gamma(\lambda)=-\ln Z[\vec{0},\hat 0;\lambda]$ holds true generally, especially for $\lambda=1$.
This completes the proof.
Thus, the $\Phi$-derivable scheme obeying Eq.\ (\ref{identity1}) includes the exact theory as a limit.

\subsection{Procedure to construct $\Phi^{(n)}$\label{subsec:Proc-Phi}}

Difficulties in constructing $\Phi^{(n)}$ for BECs originate from anomalous interaction vertices of Fig.\ \ref{Fig2}(b)-(d) that emerge upon condensation,
which make the concept of ``skeleton diagrams'' introduced for normal states \cite{LW60} obscure.
To overcome them with avoiding any prejudice, inconsistency, or double counting, 
we have previously adopted the strategy of starting from the well-established normal-state Luttinger-Ward functional 
and successively incorporating contribution of all the diagrams
characteristic of BECs so that either of identities (\ref{identity1}) and (\ref{identity2}) is satisfied.
To this end, we have relaxed the conventional 2PI condition for $\Phi$ down to 1PI,
considering that $\Phi$ obeying Eqs.\ (\ref{identity1}) and (\ref{identity2}) may not be found within the 2PI requirement.

\begin{figure}[t]
        \begin{center}
                \includegraphics[width=0.8\linewidth]{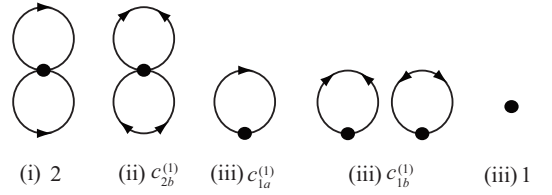}
        \end{center}
\vspace{-5mm}
        \caption{Diagrams contributing to $\Phi^{(1)}$.  A line with an arrow denotes $G$, and a line with two arrows signifies either $F$ or $\bar{F}$ as in the theory of superconductivity. \cite{AGD63}
        Here (i)-(iii) distinguish three kinds of diagrams considered at different stages of the procedure given in the second paragraph 
        of \S\ref{subsec:Proc-Phi},
and numbers and unknown variables $c^{(1)}_{\nu}$ ($\nu=2b,1a,1b$) denote relative weights of the diagrams.  
Each weight should be multiplied by $1/2\beta$ to 
        obtain the absolute weight.\label{Fig3}}
\end{figure}

\begin{figure}[b]
        \begin{center}
                \includegraphics[width=0.85\linewidth]{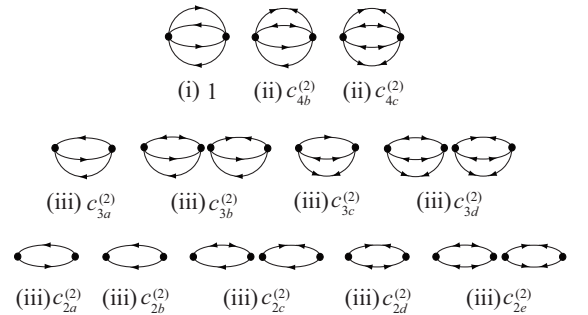}
        \end{center}
        \vspace{-5mm}
        \caption{Diagrams contributing to $\Phi^{(2)}$. Here (i)-(iii) distinguish three kinds of diagrams considered at different stages of the procedure given in the second paragraph of \S\ref{subsec:Proc-Phi}, 
        and number $1$ and unknown weights $c^{(2)}_{\nu}$ ($\nu=4b,\cdots,2e$) denote relative weights of these diagrams.  Each weight should be multiplied by $-1/2\beta$ to obtain the absolute weight.\label{Fig4}}
\end{figure}

The explicit procedure to construct $\Phi$ is summarized in terms of functions in Eq.\ (\ref{GFbF}) as (i)-(iv) below.
See Figs.\ \ref{Fig3} and \ref{Fig4} for relevant diagrams of $n=1$ and $2$, respectively.

\begin{itemize}
\vspace{-2mm}
\item[(i)] Draw all the normal-state diagrams contributing to $\Phi^{(n)}$, i.e., diagrams that appear in the Luttinger-Ward functional. \cite{LW60}
With each such diagram, associate the known weight of the normal state.

\vspace{-2mm}
\item[(ii)] Draw all the distinct diagrams obtained from those of (i) 
by successively changing directions of a pair of incoming and outgoing arrows at each vertex.
This enumerates all the processes where $F$ or $\bar{F}$ characteristic of condensation is relevant in place of $G$.
With each such diagram,  associate an unknown weight $c_\nu$.

\vspace{-2mm}
\item[(iii)] Draw all the distinct 1PI diagrams obtained from those 
of (i) and (ii) by successively removing a line, i.e., Green's function. 
This exhausts processes where the condensate wave function participates explicitly.
The 1PI condition guarantees that the self-energies obtained by Eq.\ (\ref{Sigma-Phi1}) are composed of connected diagrams.
Associate an unknown weight with each such diagram, except the one consisting only of a single vertex in the first order, i.e., the rightmost diagram in Fig.\ \ref{Fig3},
for which the weight is easily identified to be $1/2\beta$.  Indeed, the latter represents the term obtained from Eq.\ (\ref{S_int}) by replacing every field operator by its expectation value, 
i.e., the condensate wave function.

\vspace{-2mm}
\item[(iv)] Determine the unknown weights of (ii) and (iii) by requiring that either Eq.\ (\ref{identity1a}) or (\ref{identity2a}) be satisfied.
\end{itemize}

\vspace{-2mm}
Now, we apply the above procedure to constructing $\Phi^{(1)}$. Its diagrams are enumerated in Fig.\ \ref{Fig3}.
The corresponding analytic expression is given by
\begin{align}
\Phi^{(1)}\!=&\,\frac{1}{2\beta}\int dx'\int dy'\int dx\int dy \,V_{\rm s}(x'y',xy)\nonumber \\
&\times\bigl\{2G(x,x')G(y,y')+c_{2b}^{(1)}F(x,y)\bar F(x',y')
\nonumber \\
&+c_{1a}^{(1)}G(x,x')\Psi(y)\bar\Psi(y')
+c_{1b}^{(1)}\bigl[F(x,y)\bar\Psi(x')\bar\Psi(y')
\nonumber \\
&+\bar F(x',y')\Psi(x)\Psi(y)\bigr]
+\bar\Psi(x')\bar\Psi(y')\Psi(x)\Psi(y)\bigr\} .\nonumber
\end{align}
Introducing the rule of associating non-primed (primed) arguments with $\psi$ ($\bar\psi$),
we may express this $\Phi^{(1)}$ concisely as
\begin{subequations}
\label{Phi^(1)-old}
\begin{align}
\Phi^{(1)}\!=&\, \frac{1}{2\beta}{\rm Tr}\,V_{\rm s}\bigl[ 2GG+c_{2b}^{(1)}F\bar F+c_{1a}^{(1)}G\Psi\bar\Psi
\nonumber \\
&\,+c_{1b}^{(1)}\bigl(F\bar\Psi\bar\Psi+\bar F\Psi\Psi\bigr)+\bar\Psi\bar\Psi\Psi\Psi\bigr],
\end{align}
where $G G$, $F\bar F$, etc., are matrices with elements 
$G(x,y')G(y,x')$, $F(x,y)\bar F(x',y')$, etc.
We now require that Eq.\ (\ref{identity1a}) be satisfied, whose differentiations graphically correspond 
to removing a line of $G$, $F$, and $\bar F$ from Fig.\ \ref{Fig3} in all possible ways, respectively.
In this process, prefactors of Fig.\ \ref{Fig3}(i) and (ii) with $2n$ ($n=1$) Green's function lines vanish identically from the equation. 
This cancellation is characteristic of those diagrams with no condensate wave function
and holds true order by order in Eq.\ (\ref{identity1a}). 
Hence, Eq.\ (\ref{identity1a}) for $n=1$ is reduced to coupled algebraic equations for prefactors of latter three diagrams in Fig.\ \ref{Fig3},
which read $c^{(1)}_{1a}=\frac{1}{2}(-2\times 2+c^{(1)}_{1a})$, $c^{(1)}_{1b}=\frac{1}{2}(-c^{(1)}_{2b}+c^{(1)}_{1b})$,
$1=\frac{1}{2}(-c^{(1)}_{1a}-2c^{(1)}_{1b})$, respectively.
Solving them, we obtain
\begin{align}
c^{(1)}_{2b}=-1,\hspace{5mm}c^{(1)}_{1a}=-4,\hspace{5mm}c^{(1)}_{1b}=1.
\label{c^(1)}
\end{align}
\end{subequations}
It turns out that another identity (\ref{identity2a}) with $\Phi\rightarrow \Phi^{(1)}$ also yields Eq.\ (\ref{c^(1)}).\cite{Kita09}
Thus, weights $c^{(1)}_{\nu}$ have been determined uniquely based on Eqs.\  (\ref{identity1a}) and (\ref{identity2a}).

\begin{figure}[t]
        \begin{center}
                \includegraphics[width=0.4\linewidth]{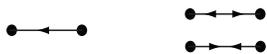}
        \end{center}
        \vspace{-5mm}
        \caption{Two kinds of extra diagrams relevant to the calculation of Eq.\ (\ref{identity1a}) for $n=2$.}
        \label{Fig5}
\end{figure}

Diagrams for $\Phi^{(2)}$ are enumerated in Fig.\ \ref{Fig4}. 
Subsequently, we require that Eq.\ (\ref{identity1a}) for $n=2$ be satisfied,
whose differentiations can also be performed graphically.
The condition yields coupled algebraic equations originating from the prefactor for each diagram in Figs.\ \ref{Fig4}
and \ref{Fig5}.
Those for the first row of Fig.\ \ref{Fig4} with $2n$ ($n=2$) lines vanish identically, as mentioned earlier.
On the other hand, equations for the second and third rows in Fig.\ \ref{Fig4} and 
those for Fig.\ \ref{Fig5} are obtained as\cite{Kita09}
\begin{align}
0=&\,c^{(2)}_{3a}+4=c^{(2)}_{3b}+ c^{(2)}_{4b}
=c^{(2)}_{3c}+2c^{(2)}_{4b}=c^{(2)}_{3d}+ 2c^{(2)}_{4c},
\nonumber \\
0=&\,c^{(2)}_{2a}+c^{(2)}_{3a}+c^{(2)}_{3b}=2c^{(2)}_{2b}+c^{(2)}_{3a}=2c^{(2)}_{2c}+c^{(2)}_{3c}+2c^{(2)}_{3b}
\nonumber \\
=&\, 2c^{(2)}_{2d}+c^{(2)}_{3c}+4c^{(2)}_{3d}=2c^{(2)}_{2e}+c^{(2)}_{3d},
\nonumber \\
0=&\,c^{(2)}_{2a}+c^{(2)}_{2b}+c^{(2)}_{2c}=
c^{(2)}_{2c}+c^{(2)}_{2d}+2c^{(2)}_{2e},
\nonumber
\end{align}
respectively.
Solving them, we obtain
\begin{align} 
&c^{(2)}_{4b}=-2, \hspace{3mm} c^{(2)}_{4c}=1,
\nonumber \\
& c^{(2)}_{3a}=-4,\hspace{3mm}c^{(2)}_{3b}=2,\hspace{3mm}  c^{(2)}_{3c}=4,\hspace{3mm}  
c^{(2)}_{3d}=-2,
\label{c^(2)}\\
&
c^{(2)}_{2a}=c^{(2)}_{2b}=2,\hspace{3mm} c^{(2)}_{2c}=-4,\hspace{3mm} c^{(2)}_{2d}=2,\hspace{3mm} c^{(2)}_{2e}=1.
\nonumber 
\end{align}
It turns out that another identity (\ref{identity2a}) with $\Phi\rightarrow \Phi^{(2)}$ also yields Eq.\ (\ref{c^(2)}).\cite{Kita09}

Thus, weights $c^{(2)}_{\nu}$ have been determined uniquely so as to satisfy both Eqs.\  (\ref{identity1a}) and (\ref{identity2a}).
However, this has been possible only by relaxing the 2PI condition for $\Phi$ down to 1PI.
Indeed, one may check easily by repeating the above calculation that $\Phi^{(2)}$ obeying either Eq.\  (\ref{identity1a}) or  (\ref{identity2a})
cannot be found within the 2PI requirement of retaining only diagrams in the first and second rows of Fig.\ \ref{Fig4}.

\subsection{Discrepancy with the CJT formalism\label{subsec:CJT}}

Thus, the above analysis has shown that the $\Phi$ functional for BECs satisfying Eqs.\ (\ref{identity1a}) and (\ref{identity2a}) 
may only be constructed by including 1PI diagrams that lie outside the 2PI category.
This conclusion has been reached based on a single requirement that $\Phi$ be expanded in terms of the interaction as Eq.\ (\ref{Phi-exp})
like the Luttinger-Ward functional.
The condition also has been crucial for proving convergence of the series to the exact action, as given below Eq.\ (\ref{identity2a}).
However, the resultant $\Phi$ apparently contradicts the one obtained by CJT,\cite{CJT74} 
which consists of 2PI diagrams even for spontaneous broken-symmetry phases of $\langle \psi\rangle\neq 0$.

The proof by CJT 
for $\Phi$ being 2PI is based on the correspondence of Eq.\ (2.19) for $\Gamma(\phi,G)$
to Eq.\ (2.10) for $\Gamma(0,G)$, \cite{CJT74}
with $\phi\rightarrow \vec\Psi$ and $ G\rightarrow  \hat G$ from their notation to ours.
However, it may not be entirely clear in the context of Bose-Einstein condensation.
First, $\Gamma(0,G)$ is relevant to normal states,
so that the external source $J^0$ for $\phi=0$ in Eq.\ (2.10) is necessarily equal to zero, i.e.,
the tadpole vertex of Fig.\ \ref{Fig2}(d) is absent in $\Gamma(0,G)$.
Thus, $\Gamma(0,G)$ is exactly the Luttinger-Ward functional of Fig.\ \ref{Fig1}
that is composed only of the vertex of Fig.\ \ref{Fig2}(a).
On the other hand, $\Gamma(\phi,G)$ contains vertices of Fig.\ \ref{Fig2}(b)-(d) inherent in BECs besides 
the classical one of Fig.\ \ref{Fig2}(e).
The vertex of Fig.\ \ref{Fig2}(c) has been removed by CJT to introduce another propagator ${\cal D}$ with it, which
reads in terms of Eqs.\ (\ref{hatG_0^-1}) and (\ref{Gamma^(0)}) in the present notation as
\begin{align}
\hat{\cal D}^{-1}(x,y)\equiv& \hat{G}_0^{-1}(x,y)-2V_{\rm s}(xy_1,yx_1)\Psi(x_1)\bar\Psi(y_1)\hat\sigma_3
\nonumber \\
&-V_{\rm s}(xy,x_1y_1)\Psi(x_1)\Psi(y_1)\frac{\hat\sigma_1+i\hat\sigma_2}{2}
\nonumber \\
&+V_{\rm s}(x_1y_1,xy)\bar\Psi(x_1)\bar\Psi(y_1)\frac{\hat\sigma_1-i\hat\sigma_2}{2},
\nonumber 
\end{align}
where  integrations over repeated arguments $(x_1,y_1)$ are implied.
Thus, part of the interaction effects have been incorporated into the ``bare'' propagator ${\cal D}$.

However, it is not clear whether their 2PI series for $\phi\!\neq \! 0$ really converges to the exact $\Phi$ 
when collected up to the infinite order, due  to the asymmetric treatment of interaction vertices in Fig.\ \ref{Fig2}
as noted above.
To be more specific, it does not obey Eq.\ (\ref{identity1}) at each order
that has been crucial in proving the convergence for normal states\cite{LW60} and also for BECs as 
given below the paragraph of Eq.\ (\ref{identity2a}).
Thus, the 2PI series may contain some over- or undercounting in the process of renormalization.
Whether it converges to the exact action or not remains to be established.
In this context, the CJT formalism has a difficulty 
that one may not find any approximate $\Phi$ that satisfies Eq.\ (\ref{DB-GP}), 
as discussed by van Hees and Knoll, \cite{HK02}
who thereby proposed a further approximation $\tilde\Gamma[\vec{\Psi}]\equiv \Gamma[\vec\Psi,\hat G[\vec\Psi]]$
to meet Eq.\ (\ref{DB-GP2}) alone.

\section{Obtaining $\Phi$ more concisely\label{sec3}}

In this section, we will show that $\Phi[\vec{\Psi},\hat{G}]$ may be simplified further to a functional of $\hat{\cal G}$  
alone defined by Eq.\ (\ref{calG}). 
Given this is the case, we realize by noting $\delta \Phi^{(n)}[\hat{\cal G}]/\delta\hat{G}=\delta \Phi^{(n)}[\hat{\cal G}]/\delta\hat{\cal G}$ that
Eq.\ (\ref{identity1}) implies a manifest fact that every term in $\Phi^{(n)}$ is composed of $2n$ products of ${\cal G}_{ij}$.
In addition, $\Phi[\hat{\cal G}]$ automatically satisfies Goldstone's theorem (I) given by Eq.\ (\ref{Sigma-Phi2}).
This is shown by using Eqs.\  (\ref{calG}) and (\ref{Sigma-Phi1}), $\Psi^i (x)=\Psi_{3-i}(x)$, and $\Sigma_{ij}(x,y)\!=\!(-1)^{i+j-1}\Sigma_{3-j,3-i}(y,x)$ as
\begin{align}
\beta\frac{\delta \Phi[\hat{\cal G}]}{\delta \Psi^{i}(x)}=&\,\sum_{j,k}\int d^4 y \int d^4 z 
\,\beta\frac{\delta \Phi[\hat{\cal G}]}{\delta {\cal G}_{jk}(y,z)}\frac{\delta {\cal G}_{jk}(y,z)}{\delta \Psi^{i}(x)}
\nonumber \\=&\,\sum_{j}\int d^4 y\, \Sigma_{ij}(x,y)(-1)^{j-1}\Psi_j(y).
\end{align}
Finally, a couple of requirements that (a) $\Phi$ be reduced to the Luttinger-Ward functional in the normal-state limit and 
(b) $\Phi$ be 1PI  will be shown to determine $\Phi[\hat{\cal G}]$ uniquely.

As a preliminary, let us define four functions in terms of ${\cal G}_{ij}$ in Eq.\ (\ref{calG}) by
\begin{subequations}
\label{calGFbF}
\begin{align}
{\cal G}(x,y') \equiv &\, \bar{\cal G}(y',x)\equiv \frac{{\cal G}_{11}(x,y')-{\cal G}_{22}(y',x)}{2}
\nonumber \\
=&\,G(x,y')-\Psi(x)\bar\Psi(y'),
\label{calG2}\\
{\cal F}(x,y)\equiv&\,\frac{{\cal G}_{12}(x,y)+{\cal G}_{12}(y,x)}{2}
\nonumber \\
=&\,F(x,y)-\Psi(x)\Psi(y),
\label{calF}\\
\bar {\cal F}(x',y')\equiv &\,-\frac{{\cal G}_{21}(x',y')+{\cal G}_{21}(y',x')}{2}
\nonumber \\
=&\,\bar F(x',y')-\bar\Psi(x')\bar\Psi(y'),
\label{calbF}
\end{align}
\end{subequations}
in exactly the same way as Eq.\ (\ref{GFbF}).
With these functions, an alternative procedure to construct $\Phi^{(n)}$ is summarized as follows:

\begin{itemize}
\vspace{-2mm}
\item[(i)] Draw all the $n$th-order diagrams of the Luttinger-Ward functional.
For each line with an arrow, associate $G$.
Identify the weight $w_j$ for each of them based on the normal-state Feynman rules. \cite{LW60}

\vspace{-2mm}
\item[(ii)] Add all the  distinct ``anomalous'' diagrams  characteristic of ODLRO that are
obtained from those of (i) 
by successively changing directions of a pair of incoming and outgoing arrows at each vertex.
For each line with an arrow (two arrows), associate $G$ ($F$ or $\bar F$).
This exhausts processes where $F$ or $\bar F$ characteristic of condensation is relevant in place of $G$.
With each such diagram,  attach an unknown weight $c_j$.

\vspace{-2mm}
\item[(iii)] Write down $\Phi^{(n)}$ based on diagrams of (i) and (ii) with replacement $(G,F,\bar F)\rightarrow({\cal G},{\cal F},\bar{\cal F})$.

\vspace{-2mm}
\item[(iv)] Determine the unknown weights of (ii) by requiring that $\Phi^{(n)}$ for $n\geq 2$ be 1PI,
and  $\Phi^{(1)}$ reproduce the classical diagram of Fig.\ \ref{Fig2}(e) with the correct weight. 
\end{itemize}

\vspace{-2mm}
Deferring detailed consideration until $\Phi^{(3)}$, 
we first present results for $n=1,2$ obtained from the above procedure:  
\begin{equation}
\Phi^{(1)}= \frac{1}{2\beta}{\rm Tr}\,V_{\rm s}\!\left(2{\cal G}{\cal G}-{\cal F}\bar{\cal F}\right),
\end{equation}
\begin{equation}
\Phi^{(2)}
=-\frac{1}{2\beta}{\rm Tr}\bigl(V_{\rm s}{\cal G}\bar{\cal G}V_{\rm s}{\cal G}\bar{\cal G}
\!-\! 2V_{\rm s}{\cal G}\bar{\cal G}V_{\rm s}{\cal F}\bar{\cal F}
\!+\! V_{\rm s}{\cal F}\bar{\cal F}V_{\rm s}{\cal F}\bar{\cal F}\bigr).
\end{equation}
It is straightforward to see that substitution of Eq.\ (\ref{calGFbF}) into these expressions reproduces
Eq.\ (\ref{Phi^(1)-old}) for $\Phi^{(1)}$ and weights of Eq.\ (\ref{c^(2)}) for $\Phi^{(2)}$.
Note that diagrams needed here are the first two in Fig.\ \ref{Fig3} for $\Phi^{(1)}$, and those of the first row
in Fig.\ \ref{Fig4} for $\Phi^{(2)}$.

\begin{figure}[t]
\begin{center}
\includegraphics[width=0.95\linewidth]{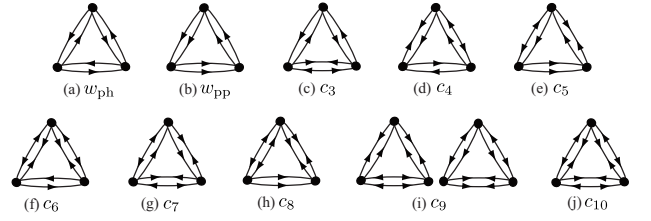}
\end{center}
\vspace{-5mm}
\caption{Distinct diagrams for $\Phi^{(3)}$ with six $G_{ij}$ lines. \label{Fig6}}
\end{figure}
\begin{figure}[b]
\begin{center}
\includegraphics[width=0.95\linewidth]{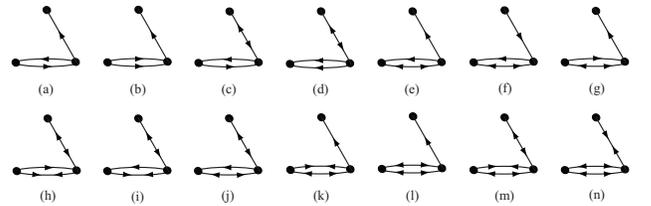}
\end{center}
\vspace{-5mm}
\caption{Leading non-1PI diagrams for $\Phi^{(3)}$ whose contribution should vanish. 
Each diagram has its partner obtained by inverting all the arrows simultaneously. \label{Fig7}}
\end{figure}
\begin{figure*}[t]
\begin{center}
\includegraphics[width=0.95\linewidth]{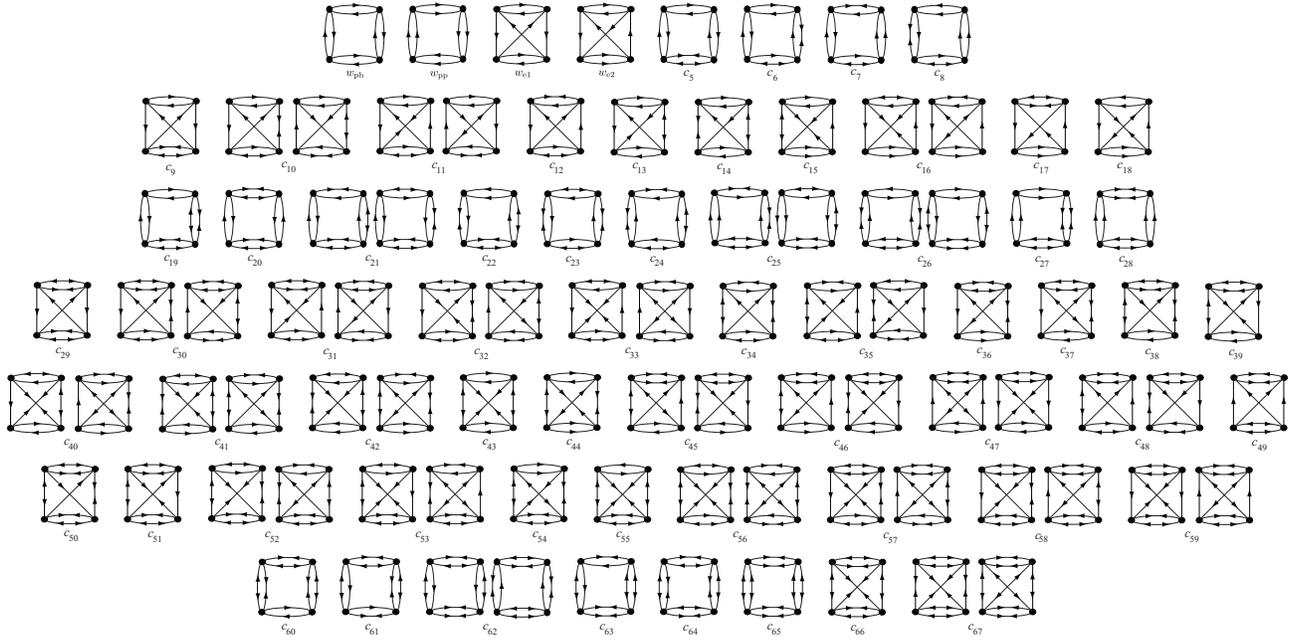}
\end{center}
\vspace{-5mm}
\caption{Diagrams for $\Phi^{(4)}$ with eight $G_{ij}$ lines. The first four survive in the normal state, whose relative weights are found to be $(w_{\rm ph},w_{\rm pp},w_{\rm e1},w_{\rm e2})=(8,1,16,16)$ with 
the common factor $-(4\beta)^{-1}$. 
The 1PI requirement for $\Phi^{(4)}[\hat{\cal G}]$ determines 
weights $c_j$ ($j=5,\cdots,67$) of the anomalous diagrams uniquely
as $4w_{\rm ph}=c_{19}= 2c_{20}=4c_{64}$, $8w_{\rm pp}=-c_{6}=c_{7}=-c_{8}=-c_{12}=c_{16}=-2c_{18}=-c_{21} = c_{22}=c_{23}
=c_{24}=-c_{25}=2c_{26} =2c_{27}=4c_{28}=-c_{32}=-2c_{33}=-c_{34}
=-c_{35}=c_{42}=-2c_{46}=c_{48}=-4c_{49}=-2c_{56}=c_{57}
= -c_{61}=2c_{62}=-c_{63}=8c_{65}=-8c_{67}$, $4w_{\rm e1}=c_{39} =4c_{43}=2c_{50}$,
$2w_{\rm e2}= -c_{11}=-c_{30}=c_{36}=-c_{41}=2c_{44}=2c_{45}=-c_{53}=2c_{59}$,
$-4w_{\rm ph}+4w_{\rm pp}= c_{5}=c_{60}$, 
$8w_{\rm pp}+2w_{\rm e2}=c_{10}=-c_{15}=-2c_{17}=c_{31} =c_{40}=-2c_{47}=c_{52}=-2c_{54}=-2c_{58}$,
$-8w_{\rm pp}-4w_{\rm e1}-2w_{\rm e2}=2c_{9}=c_{13}=c_{38}=c_{51}=2c_{55}$,
$8w_{\rm pp}+4w_{\rm e1}+4w_{\rm e2}=2c_{14}=4c_{29}=c_{37}=4c_{66}$.
\label{Fig8}}
\end{figure*}

Now, we consider $\Phi^{(3)}$ in detail. 
Here, distinct normal-state diagrams in process 
(i) above are the particle-hole and particle-particle bubble diagrams of Fig.\ \ref{Fig6}(a) and (b),  respectively,
where a line with an arrow denotes $G$.
The normal-state Feynman rules \cite{LW60,Kita09} enable us to identify their relative weights 
unambiguously as $(w_{\rm ph},w_{\rm pp})\!=\!(4,1)$ with a common factor $(3\beta)^{-1}$.
Process (ii) yields diagrams (c)-(j) of Fig.\ \ref{Fig6}, where a line with a pair of arrows toward (from) vertices denotes $F$ ($\bar F$).
Following process (iii) above, we express $\Phi^{(3)}$ analytically as
\begin{subequations}
\label{Phi^(3)}
\begin{align}
\Phi^{(3)}
=&\,\frac{1}{3\beta}{\rm Tr}\,[w_{\rm ph}V_{\rm s}{\cal G}\bar{\cal G}V_{\rm s}{\cal G}\bar{\cal G}V_{\rm s}{\cal G}\bar{\cal G}
+w_{\rm pp}V_{\rm s}{\cal G}{\cal G}V_{\rm s}{\cal G}{\cal G}V_{\rm s}{\cal G}{\cal G}
\nonumber \\
&\,
+c_{3}V_{\rm s}{\cal G}\bar{\cal G}V_{\rm s}{\cal G}\bar{\cal G}V_{\rm s}{\cal F}\bar{\cal F}+c_{4}V_{\rm s}{\cal G}\bar{\cal F}V_{\rm s}{\cal G}{\cal F}V_{\rm s}{\cal G}\bar{\cal G}
\nonumber \\
&\,
+c_{5}V_{\rm s}{\cal G}{\cal F}V_{\rm s}{\cal G}\bar{\cal F}V_{\rm s}{\cal G}{\cal G}+c_{6}V_{\rm s}{\cal F}\bar{\cal F}V_{\rm s}{\cal F}\bar{\cal F}V_{\rm s}{\cal G}\bar{\cal G}
\nonumber \\
&\,
+c_{7}V_{\rm s}\bar{\cal G}{\cal F}V_{\rm s}\bar{\cal G}\bar{\cal F}V_{\rm s}{\cal F}\bar{\cal F}+c_{8}V_{\rm s}{\cal F}{\cal F}V_{\rm s}\bar{\cal F}\bar{\cal F}V_{\rm s}{\cal G}{\cal G}
\nonumber \\
&\,
+c_{9}(V_{\rm s}\bar{\cal G}\bar{\cal F}V_{\rm s}{\cal G}\bar{\cal F}V_{\rm s}{\cal F}{\cal F}+V_{\rm s}{\cal G}{\cal F}V_{\rm s}\bar{\cal G}{\cal F}V_{\rm s}\bar{\cal F}\bar{\cal F})
\nonumber \\
&\,
+c_{10}V_{\rm s}{\cal F}\bar{\cal F}V_{\rm s}{\cal F}\bar{\cal F}V_{\rm s}{\cal F}\bar{\cal F}
\bigr] ,
\label{Phi^(3)a}
\end{align}
where ${\cal G}\bar{\cal G}$, ${\cal F}\bar{\cal F}$, etc., are matrices with elements ${\cal G}(x,y')\bar{\cal G}(x',y)$, ${\cal F}(x,y)\bar{\cal F}(x',y')$, etc.,
with ${\cal G}\bar{\cal G}$ and ${\cal G}{\cal G}$ denoting the particle-hole and particle-particle bubbles, respectively.
This $\Phi^{(3)}$ generally contains non-1PI diagrams due to the contribution of the condensate wave functions in Eq.\ (\ref{calGFbF}).
The leading ones among them are those of Fig.\ \ref{Fig7} with three $G_{ij}$ lines.
Following process (iv), we require that their contribution vanish identically.
Figure \ref{Fig7}(a), for example, is derivable from Fig.\ \ref{Fig6}(a), (c), and (d) by removing three lines adequately, and
numbers of the combinations are easily identified as 6, 2, and 1, respectively.
Thus, the requirement that Fig.\ \ref{Fig7}(a) vanish yields $6w_{\rm ph}+2c_{3}+c_{4}=0$.
The same consideration for every diagram of Fig.\ \ref{Fig7} provides:
(a) $6w_{\rm ph}+2c_{3}+c_{4}=0$, (b) $6w_{\rm pp}+c_{5}=0$, (c) $2c_{3}+c_{4}+2c_{6}=0$,
(d) $c_{5}+2c_{8}=0$, (e) $c_{4}+2c_{9}=0$, (f) $c_{4}+c_{5}=0$, (g) $c_{4}+2c_{5}+c_{7}=0$, (h) $c_{5}+c_{7}=0$, 
(i) $c_{4}+c_{7}+4c_{9}=0$, (j) $c_{7}+2c_{9}=0$, (k) $2c_{3}+2c_{6}+c_{7}=0$, (l) $2c_{8}+2c_{9}=0$, 
(m) $2c_{6}+c_{7}+6c_{10}=0$, (n) $2c_{8}+2c_{9}=0$.
They can be solved uniquely
in terms of normal-state weights $(w_{\rm ph},w_{\rm pp})=(4,1)$ as
\begin{align}
& c_{3}=3c_{10}=-3(w_{\rm ph}+w_{\rm pp}) ,\nonumber \\
& c_{4}=-c_{5}=c_{7}=2c_{8}=-2c_{9}=6w_{\rm pp}, \label{Phi^(3)-sol} \\
& c_{6}=3w_{\rm ph} .\nonumber
\end{align}
\end{subequations}
The solution also removes all the other non-1PI diagrams with less than three $G_{ij}$ lines from $\Phi^{(3)}$, as may be confirmed easily.
Expression (\ref{Phi^(3)}) with $(w_{\rm ph},w_{\rm pp})\!=\!(4,1)$ coincides exactly with that obtained previously in terms of $\hat G$ and $\vec\Psi$;
see Appendix B of ref.\ \onlinecite{Kita09}.
Looking back at Eq.\ (\ref{Phi^(3)-sol}), we also realize that each normal-state diagram of Fig.\ \ref{Fig6}(a) and (b) 
can independently be a source 
of an approximate $\Phi$ for BECs that satisfies both Noether's theorem and Goldstone's theorem (I),
i.e., we can find an approximate $\Phi$ from Fig.\ \ref{Fig6}(a) alone by setting $(w_{\rm ph},w_{\rm pp})\!=\!(4,0)$ or 
from Fig.\ \ref{Fig6}(b) alone by choosing $(w_{\rm ph},w_{\rm pp})\!=\!(0,1)$.

\begin{figure}[b]
\begin{center}
\includegraphics[width=0.95\linewidth]{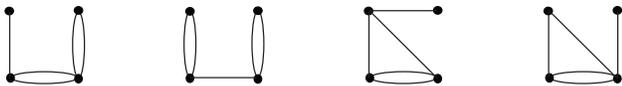}
\end{center}
\vspace{-5mm}
\caption{Leading non-1PI diagrams of the 4th order drawn without arrows.
\label{Fig9}}
\end{figure}

To confirm the validity of our procedure, we have extended our consideration to the 4th order.
The results are summarized in Fig.\ \ref{Fig8} and its caption.
The relevant normal-state diagrams in the 4th order are the first four diagrams of Fig.\ \ref{Fig8},
whose relative weights are found as $(w_{\rm ph},w_{\rm pp},w_{\rm e1},w_{\rm e2})=(8,1,16,16)$.
The requirement that ``all the leading non-1PI diagrams of Fig.\ \ref{Fig9} vanish'' has been confirmed to
determine all the weights of anomalous diagrams uniquely in terms of $(w_{\rm ph},w_{\rm pp},w_{\rm e1},w_{\rm e2})$.
The fact also implies that each of the first-four normal-state diagrams in Fig.\ \ref{Fig8}
can independently be a source for an approximate $\Phi$ that satisfies both Noether's theorem and Goldstone's theorem (I).

\section{Summary\label{sec4}}

\begin{figure}[t]
\begin{center}
\includegraphics[width=0.95\linewidth]{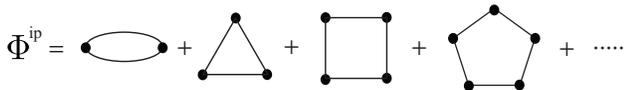}
\end{center}
\vspace{-5mm}
\caption{Approximate series $\Phi^{\rm ip}$ dominant in the dilute limit.
\label{Fig10}}
\end{figure}

We have developed a concise procedure to construct the effective action for BECs 
in such a way that both Noether's theorem and Goldstone's theorem (I) are satisfied at each order of a power series in terms of the interaction.  
It is found that every normal-state diagram can be a source of an approximate $\Phi$ for BECs.
However, this is found possible only at the 1PI level instead of 2PI due to the anomalous structures of the bare interaction vertices as
shown in Fig.\ \ref{Fig2}.
The resultant self-energy, obtained by Eq.\ (\ref{Sigma-Phi1}), should necessarily be one-particle reducible (1PR); 
this structure has been overlooked and may change our standard understanding of BECs substantially. 
For example, leading non-2PI diagrams for $\Phi$ in the dilute limit is given by the series of Fig.\ \ref{Fig10}.
They are predicted to modify the Lee-Huang-Yang expression \cite{LHY57}  for the ground-state energy per particle into \cite{TK13}
\begin{align}
\frac{E}{N}= \frac{2\pi\hbar^2 a n}{m} \biggl[1+\biggl(\frac{128}{15\sqrt{\pi}}+\frac{16}{5}c_{{\rm ip}}\biggr) \sqrt{a^3n}\biggr],
\label{LHY-E}
\end{align}
where $a$ and $n$ are the $s$-wave scattering length and particle density, respectively,
and $c_{\rm ip}=O(1)$ is an additional constant due to $\Phi^{\rm ip}$.
Moreover, this contribution is expected to change the nature of poles of $\hat{G}$, which dominate thermodynamic properties of dilute BECs,
from the Bogoliubov mode with an infinite lifetime \cite{Bogoliubov47}
 into a bubbling mode with a large decay rate proportional to $a$, \cite{TK14}
instead of $a^2$ for the normal state.
However, the fact does not contradict ``Goldstone's theorem (II)'' from the second proof based on the commutation relation,\cite{Weinberg96,GSW62} which
predicts a gapless mode with an infinite lifetime for homogeneous systems. 
As shown previously, \cite{Kita11} Goldstone's theorem (II) is relevant to three-point functions for BECs sharing poles with four-point functions,
where the 1PR structure cancels out to yield an infinite lifetime for the collective excitations. Thus, their poles are distinct from those of $\hat G$. \cite{Kita10b}
The fact illustrates that the contents of the two proofs \cite{Weinberg96,GSW62} are not identical in general and should be distinguished clearly 
as ``Goldstone's theorem (I)'' and ``Goldstone's theorem (II).'' \cite{Kita11}

\end{document}